\documentclass[prx,twocolumn,superscriptaddress,showpacs,amsmath,amstex,amssymb,citeautoscript,10pt]{revtex4-2}
\usepackage{dcolumn}% Align table columns on decimal point
\usepackage{bm}% bold math
\usepackage[T1]{fontenc}
\usepackage{braket}
\usepackage{graphicx}
\usepackage{xcolor}
\usepackage{lipsum}
\usepackage[normalem]{ulem}
\usepackage{amsmath}
\usepackage{amsfonts}
\usepackage{amssymb} 
\usepackage{color}
\usepackage[percent]{overpic}
\usepackage{soul} 
\usepackage{amssymb}
\usepackage{wasysym}
\usepackage{dsfont}
\usepackage{float}
\usepackage[mathscr]{euscript}
\usepackage{ifthen}
\usepackage{csquotes}
\usepackage{appendix}
\usepackage[bottom]{footmisc}
\usepackage{verbatim}
\usepackage{multirow}
\usepackage{mathrsfs}
\usepackage[scr=esstix,cal=boondox]{mathalfa} 
\usepackage{CJKutf8}
\usepackage{lineno}
\renewcommand{\vec}[1]{\mathbf{{#1}}}
\DeclareMathSymbol{\cdot}{\mathord}{symbols}{"01} %makes cdot spacing closer together
\begin{document}

\title{Coherent and incoherent light scattering by single-atom wavepackets}

\author{{Vitaly Fedoseev}}
\altaffiliation{These authors contributed equally to this work.}
\affiliation{Department of Physics, Massachusetts Institute of Technology, Cambridge, MA 02139, USA}
\affiliation{Research Laboratory of Electronics, Massachusetts Institute of Technology, Cambridge, MA 02139, USA}
\affiliation{MIT-Harvard Center for Ultracold Atoms, Cambridge, MA, USA}

\author{Hanzhen Lin
(\begin{CJK*}{UTF8}{gbsn}林翰桢\end{CJK*})}
 %\author{Hanzhen Lin}
 \altaffiliation{These authors contributed equally to this work.}
\affiliation{Department of Physics, Massachusetts Institute of Technology, Cambridge, MA 02139, USA}
\affiliation{Research Laboratory of Electronics, Massachusetts Institute of Technology, Cambridge, MA 02139, USA}
\affiliation{MIT-Harvard Center for Ultracold Atoms, Cambridge, MA, USA}

\author{{Yu-Kun Lu}}
\altaffiliation{These authors contributed equally to this work.}
\affiliation{Department of Physics, Massachusetts Institute of Technology, Cambridge, MA 02139, USA}
\affiliation{Research Laboratory of Electronics, Massachusetts Institute of Technology, Cambridge, MA 02139, USA}
\affiliation{MIT-Harvard Center for Ultracold Atoms, Cambridge, MA, USA}

\author{Yoo Kyung Lee}
\altaffiliation{These authors contributed equally to this work.}
\affiliation{Department of Physics, Massachusetts Institute of Technology, Cambridge, MA 02139, USA}
\affiliation{Research Laboratory of Electronics, Massachusetts Institute of Technology, Cambridge, MA 02139, USA}
\affiliation{MIT-Harvard Center for Ultracold Atoms, Cambridge, MA, USA}

\author{{Jiahao Lyu}}
\altaffiliation{These authors contributed equally to this work.}
\affiliation{Department of Physics, Massachusetts Institute of Technology, Cambridge, MA 02139, USA}
\affiliation{Research Laboratory of Electronics, Massachusetts Institute of Technology, Cambridge, MA 02139, USA}
\affiliation{MIT-Harvard Center for Ultracold Atoms, Cambridge, MA, USA}

\author{Wolfgang Ketterle}
\affiliation{Department of Physics, Massachusetts Institute of Technology, Cambridge, MA 02139, USA}
\affiliation{Research Laboratory of Electronics, Massachusetts Institute of Technology, Cambridge, MA 02139, USA}
\affiliation{MIT-Harvard Center for Ultracold Atoms, Cambridge, MA, USA}

\begin{abstract}
We study light scattering of single atoms in free space and discuss the results in terms of atom-photon entanglement and which-way information. 
Using ultracold atoms released from an optical lattice, we realize a Gedanken experiment which interferes single photons scattering off of Heisenberg uncertainty-limited wavepackets. 
We unify the free-space and trapped-atom pictures by measuring the light scattered during wavepacket expansion and show the coherence properties of the scattered light is independent of the presence of the trap.
Our experiment demonstrates the potential of using atomic Mott insulators to create single-atom wavepackets for fundamental studies.
\end{abstract}

\maketitle

\textit{Introduction}.---
The scattering of light, even by single atoms, involves rich physics. 
Photons can be scattered elastically or inelastically (i.e., with the same or a different frequency), and coherently or incoherently (i.e., with a well-defined or random phase). 
Usually, only light which is scattered elastically and coherently can interfere. 
This can be observed for single atoms by homodyning the scattered light with the exciting laser beam, for two atoms through a Young’s double slit experiment where the two slits are two single atoms, or for many atoms in a periodic array by observing diffraction peaks or Bragg scattering.

\begin{figure}[b]
    \centering
    \includegraphics[,keepaspectratio, scale=0.7]{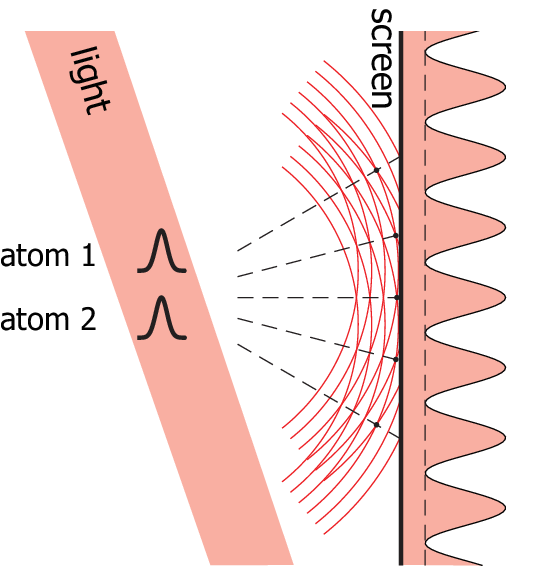}
    \caption{Light scattering from two atomic wavepackets. The light  has a coherent part which results in interference fringes, and an incoherent part resulting in a finite contrast of the fringes. The incoherent fraction originates from partial entanglement between the atoms and photons (Eq.~\ref{eq:density_matrix_light}).}
    \label{fig:experiment}
\end{figure}

We emphasize that the question of coherent and incoherent light scattering can be fully discussed for an individual atom, but without loss of generality, we focus our discussions here on two-atom interference (Fig.~\ref{fig:experiment}).
A finite contrast of the interference pattern is caused by incoherent light scattering and has been related to “which-way” information, i.e., a change of the atom’s state which provides information which atom has scattered the light. 
Seminal experiments have been carried out with trapped ions \cite{Eichmann1993,schmidt_kaler_ion_crystal_scattering_2016} and discussed in many theoretical papers \cite{Wootters1979, Scully1991, Itano1998, Utter2007,Brewer_ion_scattering_theory_1996}.
These studies demonstrated that scattering via a Raman transition to a different hyperfine state is incoherent, and also that saturating the transition leads to incoherent scattering since two photons can be scattered simultaneously, e.g., through Mollow sidebands \cite{mollow_triplet_1969}.
Some studies explored superradiance and subradiance between the two ions when they were resonantly excited \cite{Devoe1996,Eschner2001}.
Here, we focus on the simplest possible situation: an atomic two-level system, in the weak excitation limit, using far detuned light resulting in low optical density. This is the simple Rayleigh scattering limit. 

For particles in an ideal harmonic trapping potential, scattering can occur on the carrier or via sidebands. 
The relative intensity in the carrier is given by the Debye-Waller factor
$D=\exp(-\eta^2)$, where $\eta = Qx_0$ is the Lamb-Dicke parameter for momentum transfer $Q\equiv\lvert\vec{k}_{\rm in}-\vec{k}_{\rm out}\rvert$ and one-dimensional wavefunction spread $x_0$.
The idealized double slit experiment would now be described as follows: Interference is only observed for the carrier light whose frequency spectrum is a delta function at the excitation frequency. 
This line is also called the Mössbauer line, since the recoil of light scattering is not transferred to the atom, but absorbed by the lattice potential. 
Since Mössbauer scattering does not leave any which-way information, such light coherently interferes with 100\% contrast. 
On the other hand, sideband emission with relative intensity $1-D$ is shifted in frequency from the carrier (by integer multiples of the harmonic oscillator frequency) and imparts which-way information, since after the scattering event, the atom is in a different harmonic oscillator state; this corresponds to inelastic and incoherent light scattering. 
All this is well established, both experimentally for trapped ions \cite{Eichmann1993,schmidt_kaler_ion_crystal_scattering_2016} and theoretically \cite{Wootters1979, Scully1991, Itano1998, Utter2007,Brewer_ion_scattering_theory_1996}.
Other groups have studied coherent Bragg scattering from atoms trapped in optical lattices \cite{huleet_afm_order_2015,pan_afm_order_2024,junru_supersolid_stripe_2017,miyake_mott_bragg_scattering_2011} and atoms in optical cavities \cite{yan_cavity_superradiant_2023,Rempe_cavity_rayleigh_scatteirng_2010}.

Here we look at the different situation of atoms in identical Gaussian wavepackets which expand in free space, and scatter light off them with a short light pulse. 
There is no trapping potential to absorb the momentum, so strictly speaking, all light scattering is inelastic since the scattered light has a recoil shift.
However, the recoil shift is smaller than the Fourier width of the short light pulse. 
There are no harmonic oscillator states to record which-way information, and the momentum width of the localized wavepacket is much larger than the recoil.

To the best of our knowledge, coherent light scattered or emitted by freely expanding wavepackets has been observed only in two kinds of previous experiments.
One was done by Alain Aspect and collaborators \cite{alain_aspect_calcium2_recoil_interference_1982}, where the spontaneous emission from photodissociated ${\rm Ca_2}$ molecules showed temporal oscillations due to the beat note between traveling atomic fragments.
The other studies showed how coherent Bragg scattering from atoms expanding from states localized in an optical lattice changed with the duration of time-of-flight \cite{phillips_brag_scattering_lattice_1995, huleet_afm_order_2015, miyake_mott_bragg_scattering_2011}. 
In contrast, our experiment studies for the first time the (usually much weaker) incoherently scattered light which provides which-way information. We show both theoretically and experimentally that the scattered light is only partially coherent, depending on the partial entanglement between the light and atomic motion.

\textit{Theoretical description}.---
Light scattering occurs into all angles, but we can consider one scattering mode corresponding to a momentum transfer $\vec{Q}$. 
For sufficiently far-detuned light, emitted photons are not re-absorbed and do not interact with each other; therefore, each mode can be regarded independently. 
We consider two atoms with mass $m$ localised in Gaussian wavepackets with the one-dimensional root-mean-square (rms) size $x_0=\sqrt{\hbar/(2m\omega)}$ obtained by suddenly releasing them from the ground states $\ket{\alpha=0}$ of two harmonic traps at position $\vec{R}_1$ and $\vec{R}_2$ (Fig.~\ref{fig:experiment}).
The light in the considered mode is initially in the vacuum state $\ket{0}_{\rm photon}$.
Thus, the initial state for the atoms and the light is $\ket{\alpha=0}_1\ket{\alpha=0}_2\ket{0}_{\rm photon}$, where $\alpha$ is the usual notation for coherent harmonic oscillator states before trap release \cite{footnote_1d}. 
%\textcite{footnote_1d}.

\begin{figure}[t]
    \includegraphics[keepaspectratio, scale=0.75]{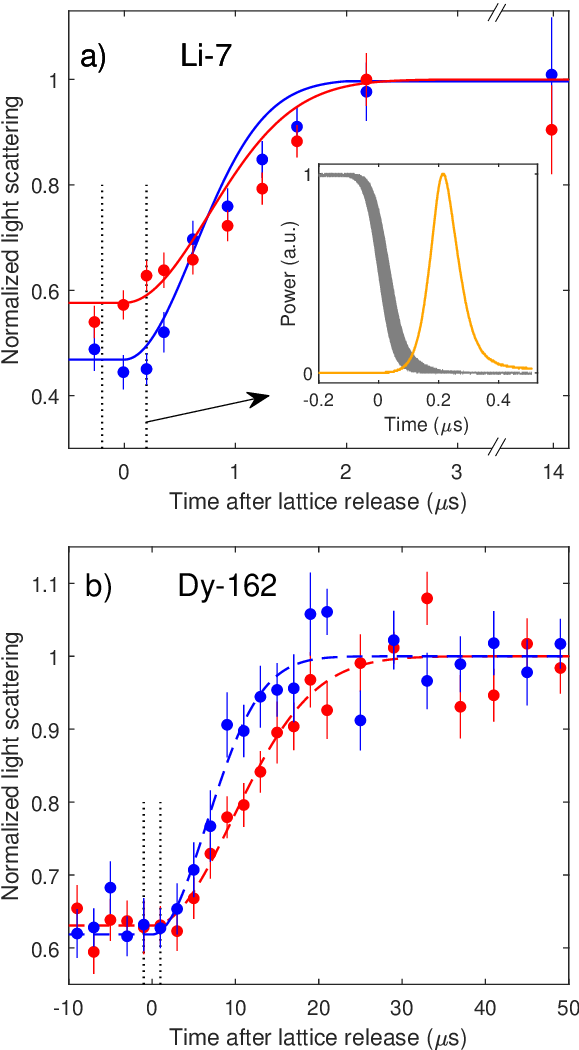}
    \caption{    
    Light scattering from expanding wavepackets. 
    Lithium (a) and dysprosium (b) atoms are prepared in a Mott insulating state in a deep (blue) or shallow (red) optical lattice. 
    Light scattering is suppressed due to the destructive interference of coherently scattered light. 
    All points before (after) the left (right) vertical dotted line indicate points taken with the lattice fully on (off). 
    The suppression is the same before and right after the switch-off, 
    confirming that the coherence properties of single atoms are independent of confinement.
    As the wavepackets expand, the scattered light becomes fully incoherent.
    Solid lines in (a) are theory with no free parameters, while the dashed lines in (b) are theory with a fraction of coherent scattering before the lattice release as a free parameter.
    Inset shows the relative timing of the probe pulse (orange) and lattices switching off (gray) corresponding to the right dotted line.
    Error bars are 1 standard error of the mean.   
    }
    \label{fig:fig2}
\end{figure}

In the weak excitation limit, there is a small probability to add a photon to the scattering mode while imparting a recoil momentum of $\hbar \vec{Q}$ to the scattering atom. 
This shifts the ground state wavepacket into the state $\ket{\beta} = \exp(i\vec{Q}\cdot\hat{\vec{R}})\ket{0}$ where $\hat{\vec{R}}$ is the displacement operator of the atom from the center of each trap and $\beta = i Q x_0$. 
The resulting total state can be described as
\begin{align}
    \ket{\Psi} &= 
    \ket{0}_1\ket{0}_2\ket{0}_{\rm photon}+ \nonumber \\
    & + \epsilon (\gamma_1 \ket{\beta}_1\ket{0}_2\ket{1}_{\rm photon} + \gamma_2 \ket{0}_1\ket{\beta}_2\ket{1}_{\rm photon}).
    \label{eq:state}
\end{align}
Here $\epsilon \ll 1$ is small parameter describing the amount of admixture while 
$\gamma_1= \exp(i\vec{Q}\cdot\vec{R}_1)$ and $\gamma_2= \exp(i\vec{Q}\cdot\vec{R}_2)$ describe the scattering phase of the atoms.
The density operator for the light in the photon mode $\rho_{\rm photon}={\rm Tr}_{\rm atoms}\left\{\ket{\Psi}\bra{\Psi}\right\}$  is obtained by performing the partial trace over the atomic states:
\begin{equation}
\begin{split}
\rho_{\rm photon} &=(\ket{0} + \epsilon\braket{0|\beta}(\gamma_1+\gamma_2)\ket{1})\\
&\times(\bra{0} + \epsilon^*\braket{\beta|0} (\gamma_1^*+\gamma_2^*)\bra{1})\\
&+2\epsilon \epsilon^*(1-\lvert\braket{\beta|0}\rvert^2)\ket{1}\bra{1}.
\end{split}
\label{eq:density_matrix_light}
\end{equation}
We identify the Debye-Waller factor as $D=\lvert\bra{\beta}0\rangle\rvert^2 = \lvert\langle \exp(i\vec{Q}\cdot\vec{R})\rangle\rvert^2=\exp(-\eta^2)$. 
The scattered light is in a statistical mixture of two states: (1) a coherent state where the photons from the two atoms interfere and which has a photon number of $\epsilon  \epsilon^* D \lvert (\gamma_1 + \gamma_2)\rvert^2$ and (2) a Fock state with one photon which has a undetermined phase and corresponds to incoherent light with probability $2 \epsilon  \epsilon^* (1 - D)$. 
For a single atom, the same treatment would give a coherent state with photon number $\epsilon  \epsilon^* D $, and a Fock state with $\epsilon  \epsilon^* (1 - D)$.
A more general treatment of quantum entanglement in photon-atom scattering can be found in refs. \cite{eberly_photon_atom_entanglement_2003,eberly_photon_atom_wavepacket_2005, Liao2005, Utter2007}. 

Our derivation shows that the fractions of coherently ($D$) and incoherently scattered light ($f_{\rm incoh}=1-D$) are the same regardless of the presence of a trapping potential. 
Therefore, recoilless scattering in a crystal (Mössbauer effect), the different frequency of sidebands, and the excitation of an excited harmonic oscillator state are not essential to the question which fraction of light scattering is coherent or incoherent. 
What matters is only the partial entanglement between light and atoms.

A classical distribution of oscillating antennas identical to the atomic density given by the square of the wavefunction would scatter light only coherently. There is no randomness which could result in light with a random phase. 
This classical model predicts the same amount of coherently scattered light, but no incoherent light.  
We can modify the classical model by assuming that for every photon scattering event, the atoms are point particles with a position randomly chosen with a weight given by the square of the atomic wavefunction. 
This model gives exactly the same amount of coherent and incoherent light scattering as the quantum treatment. 
However, it is physically wrong to assume that light scattering localizes the atom to a point source: the reason for incoherent light scattering is partial entanglement.
However, in both pictures, incoherence comes from the spatial delocalization of the wavepacket: in the classical picture it gives rise to a random phase, while in the quantum picture it results in entanglement and subsequently a Fock state with all phases.

\textit{Experiments using Mott insulators}.---
We load an optical lattice of $\sim$10,000 ultracold atoms in a singly occupied Mott insulator state and count photons far from Bragg angle to characterize the coherence properties of the scattered light. 
This many-slit generalization of Young's double-slit experiment has many advantages. 
The large number of scatterers greatly increases the signal-to-noise ratio. 
Furthermore, by looking far from the Bragg angle, our detection does not need the angular resolution $\sim1/N^{1/3}$ required to resolve the Bragg peak.
In the lattice, the atoms are in their motional ground state in Heisenberg uncertainty-limited wavepackets with $\Delta x \Delta p_x = \hbar/2$. 
The atoms are probed before, during and after the sudden switch-off of the lattice.

For an array of $N$ atoms, the normalized interference factor in Eq. \ref{eq:density_matrix_light} $\lvert (\gamma_1 + \gamma_2)\rvert^2/2$ can be generalized to the structure factor of the atom array  $S(\vec{Q}) = \lvert \Sigma_{j} e^{i\vec{Q}\cdot\vec{R}_j} \rvert^2/N$ where $j$ is a sum over all occupied sites.
For two atoms, $S(\vec{Q})$ is the standard double-slit pattern.
For a perfect crystal of infinite size, $S(\vec{Q})=0$ unless the angle of the incoming laser beam fulfills a Bragg condition [for which $S(\vec{Q})=N$].
For a cubic lattice with cubic surfaces $S(\vec{Q}) \sim 1/N$ far enough from the Bragg angles. 
This allows us to describe the normalized light scattering intensity
\begin{equation}
I(\vec{Q}) = D S(\vec{Q}) + f_{\rm incoh}.
\label{eq:eq3}
\end{equation}
Here $D$ and $f_{\rm incoh}$ are the fractions of light scattered coherently or incoherently, respectively, by a single atom.

For all geometries where $S(\vec{Q})$ is known, it is possible to distinguish coherent from incoherent scattering.
We observe scattering far from the Bragg angle, where $S(\vec{Q})$ is almost completely suppressed for a Mott insulator in a 3D rectangular lattice due to destructive interference. 
This allows us to remove the coherent part of scattering and measure that both trapped and expanding wavepackets scatter light incoherently with a normalized
intensity given by $f_{\rm incoh} = 1-D$, thus confirming the theoretical prediction of Eq.~\ref{eq:density_matrix_light}.

\textit{Results}.---
We performed similar experiments using both Li-7 and Dy-162 atoms to make use of their different properties.
The lithium setup prepares a Mott insulator of $3\times10^4$ atoms with almost exactly one atom per site.
The sample is prepared at high magnetic fields (868.8 G) in the second-lowest hyperfine state. Light is scattered using a (nearly) cycling $\sigma^-$ transition from the $\ket{J=1/2,m_J=-1/2}$ to the $\ket{J'=3/2,m_J'=-3/2}$ state at 671 nm. 
Fig.~\ref{fig:fig2} shows the main result of the paper.
Atomic wavepackets are probed before and after the switch-off of the optical lattice. 
The lattice is switched off within 0.1 $\mu$s (Fig.~\ref{fig:fig2}a inset), and the light pulses are 0.1 $\mu$s in FWHM. 
The suppression of light scattering is proportional to the fraction of coherent light scattering (Eq.~\ref{eq:eq3}). 
We note that after sufficiently long time-of-flight expansion, $D=0$ and $f_{\rm incoh}=1$. Therefore, our data for light scattering was normalized to 1 for long times when $D<0.01$.

The scattering intensity $I$ is identical before and immediately after the lattice switch-off within experimental uncertainty (Fig.~\ref{fig:fig2}a). 
This is the first major result of our experiment: the presence of a trapping potential does not alter the incoherent scattering of the light.
Furthermore, we performed a complementary experiment scattering the same number of photons with either a short (0.1 $\mu$s) or long (4 $\mu$s) pulse corresponding to Fourier widths of $2\pi\times1.6{\rm MHz}\gg\omega$ and $2\pi\times40{\rm kHz}\ll\omega$, respectively, for a trap frequency of $\omega\approx2\pi\times250$ kHz. 
The short and long pulses yielded similar scattering intensities of $I=0.48\pm0.04$ versus $I=0.52\pm0.04$.
This illustrates that the coherence properties of the scattered light are the same whether or not the sidebands are resolved, and that coherent and incoherent light can have the same frequency.

We can predict the initial scattering intensity by calculating the Debye-Waller factor $D$ and the structure factor $S(\vec{Q})$.
The deep (shallow) trap of $\omega=2\pi\times256$ kHz ($2\pi\times164$ kHz) has a ground state wavepacket of size $x_0=53$ nm ($66$ nm), corresponding to a Debye-Waller factor of 0.61 (0.49). 
We expect a structure factor $S(\vec{Q}) \approx 0.10\pm0.02$ due to two effects.
The irregularity of the approximately spherical surface of the sample contributes 0.02 (obtained from simulations) and the presence of an estimated 5--10\% holes in the Mott insulator \cite{Jepsen2020} adds $0.08\pm0.02$. 
A more detailed discussion of imperfect crystals will be presented in a forthcoming paper.
After including small corrections for the saturation parameter $s=0.02$ of the probe light [which reduces the coherent light fraction by $1/(1+s)$] and for the branching ratio of the excited state (98\% cycling), we predict normalized scattering intensities of $I=0.47\pm0.02$ (deep lattice) and $I=0.58\pm0.02$ (shallow lattice).
Our observations quantitatively agree with our theoretical description without any free parameters (Fig.~\ref{fig:fig2}a, solid lines) 

For the expanding wavepacket, the Debye-Waller factor decays as $D(t)=\exp[-Q^2x(t)^2]$ where $x(t)=x_0 \sqrt{1+(\omega t)^2}$ is the rms size of a wavepacket released from a harmonic oscillator potential \cite{Schiff1955}. Experimental observations are in semi-quantitative agreement with this prediction.
Due to available laser power, the lithium experiment was limited by the range of trap frequencies where Mott insulators could be prepared. Also, the fast expansion of light atom made it challenging to have probe laser pulses much faster than the expansion time.  We therefore performed an additional experiment with the much heavier dysprosium atom.

The Dy Mott insulator state typically has $3\times10^4$ atoms in the $\ket{J=8,m_J=-8}$ state in a distribution of Mott insulator shells with single site occupation numbers between 1 and 3. 
Light is mainly scattered using the cycling $\ket{J=8,m_J=-8}$ to $\ket{J'=9,m_J'=-9}$ transition at 626 nm. 
The structure factor $S(\vec{Q})$ for the Dy Mott insulator is $\approx 0.3$ due to a higher number of defects (holes and doublons) than in the lithium sample and was obtained as a fit parameter to the data in Fig.~\ref{fig:fig2}b. 
The dysprosium results confirm with higher time resolution that the coherence properties of the scattered light are the same before and right after switch off.  For two different trap frequencies of $2\pi\times21$ kHz and $2\pi\times43$ kHz, the temporal variation of the incoherently scattered light is in quantitative agreement with the predicted expansion of Gaussian wavepackets (Fig.~\ref{fig:fig2}b). The identical suppression of light scattering at the initial time implies that the larger Debye Waller factor of 0.87 versus 0.74 is compensated by a larger number of defects in the deeper lattice, or that the deeper lattices induce inelastic scattering of light by doublons.

\textit{Discussion}.---
Our experiment has compared three different situations for light scattering of wavepackets: Long pulses in the optical lattice (for which the sidebands are resolved) and short pulses, either before or after switch-off. 
In the long pulse experiment, which-way information is directly obtained from the probability of populating excited harmonic oscillator states, as described by Fermi's Golden rule. 
In contrast, for the short light pulse (while the lattice is on), we create a momentum-displaced wavepacket (exactly as in free space, right after the lattice is switched off) where all atoms move into the direction of the momentum transfer. 
This is how which way information is created. 
However, during a sufficiently long wait time, the coherent motion of the atoms will dephase, and we will find the atoms in a mixture of states with the same fraction of atoms in excited harmonic oscillator states as for the long-pulse experiment. 
This suggests that we could obtain which-way information for the wavepackets in free space by suddenly switching on a harmonic oscillator potential immediately after the light pulse and analyzing the populations in harmonic oscillator states. 
Although any light scattering in free space transfers recoil to the atoms, this analysis will project a fraction of the atoms onto the initial state (i.e., without recoil) corresponding to the fraction of coherently scattered light. This establishes a protocol how the entanglement and which-way information encoded in wavepackets could be measured.

Our theoretical treatment can be expanded to arbitrary wavepackets \cite{Scully1991} including thermal states.
The fraction of coherently scattered light is given by the Debye-Waller factor $\exp(-Q^2 x_0^2)$, where $x_0$ is the rms width.
It does not depend on the momentum width. If an atom at higher temperature is confined to the same size as a ground state atom in a weaker harmonic oscillator, the amount of coherently scattered light is the same, although the momentum spread of the wavepacket is larger. Similarly, as we have shown, the Debye-Waller factor decays for an expanding wavepacket although the momentum spread stays the same. If we interpret which-way information as a measurement of the momentum transfer, it is at first sight surprising that the sensitivity of this measurement does not depend on the momentum width of the atom, only the spatial width. Mathematically, it is explained by the fact that the generator for a momentum shift is the position operator.

\textit{Outlook}.---
We have shown that a single uncertainty-limited wavepacket scatters light incoherently, and related this to partial entanglement and which-way information, when two or more atoms scatter light. 
We have experimentally confirmed the prediction for the ratio of coherent to incoherent scattering. 
Our work also demonstrates that optical lattices are a powerful tool to prepare quantum controlled wavepackets of atoms and study their properties. 
Recent work
\cite{Verstraten2024} used a quantum gas microscope to study the expansion dynamics of single-atom wavepackets. 

For future work, it looks promising to apply our method of distinguishing coherent and incoherent light to wavepackets with two atoms, i.e., prepare an $n=2$ Mott insulator state in an optical lattice \cite{bec4_2022}. 
For non-interacting atoms, we expect (normalized to the number of atoms) the same amount of incoherent light scattering.  However,the dipolar interaction between an excited and ground state atom leads to a $1/R^3$ potential, which can accelerate the atoms during light scattering. 
This is a well-known loss process in magneto-optical traps called radiative escape \cite{Gallagher1989}.
For sufficiently tight confinement, this process will transfer more momentum to the atoms than the recoil momentum and will be the dominant source of which-way information and therefore reduce the coherent scattering fraction. 
Another extension would be to prepare atoms in superposition states or mixtures of spin states and study the coherence of scattered light.\\

% Data availability statement
The data that support the findings of this study are available from the corresponding author upon reasonable request.\\

% Acknowledgments
We acknowledge discussions with Peter Zoller, Serge Haroche, Eugene Polzik at the IQOQI conference in Sept. 2024. We thank Yaashnaa Singhal for experimental assistance, and Jinggang Xiang and William R. Milner for critical reading of the manuscript.
For the lithium experiment, we acknowledge support from the NSF through grant No. PHY-2208004, from the Center for Ultracold Atoms (an NSF Physics Frontiers Center) through grant No. PHY-2317134, the Army Research Office (contract No. W911NF2410218) and from the Defense Advanced Research Projects Agency (Grant No. W911NF2010090). The dysprosium experiment is supported by a Vannevar-Bush Faculty Fellowship (grant no. N00014-23-1-2873), from the Gordon and Betty Moore Foundation GBMF ID \# 12405), and DARPA (award HR0011-23-2-0038). Yoo Kyung Lee is supported in part by the National Science Foundation Graduate Research Fellowship under Grant No.1745302. Yoo Kyung Lee and Hanzhen Lin acknowledge the MathWorks Science Fellowship. Yu-Kun Lu is supported by the NTT Research Fellowship.\\

% Author contributions
All authors conceived the experiment, discussed the results and contributed to the writing of the manuscript. 
Vitaly Fedoseev, Hanzhen Lin, Yu-Kun Lu, Yoo Kyung Lee, and Jiahao Lyu performed the experiment and analyzed the data. \\

% Competing financial interests statement
The authors declare no competing financial interests.\\

\emph{Note}.---While writing the manuscript, we became aware of related simultaneous work where which-way information is studied in the context of light scattering from a single trapped atom \cite{hefei2024}.

\bibliography{mainbib}

\end{document}